\documentclass[pra,twocolumn,superscriptaddress,amsmath,amssymb]{revtex4-2}
\usepackage{amsmath}
\usepackage{amsfonts}
\usepackage{amssymb}
\usepackage{graphicx}
\usepackage{color}
\usepackage{txfonts}
\usepackage{float}
\usepackage[titletoc]{appendix}
\usepackage[colorlinks={true}]{hyperref}
\usepackage{textcase}
\hypersetup{citecolor={blue}, filecolor={blue}, linkcolor={blue}, urlcolor={blue}}

\begin{document}
\title{Chirped-pulse engineering for robust control of single-molecule orientation in a cavity}
\author{Li-Bao Fan}
\affiliation{Hunan Key Laboratory of Nanophotonics and Devices, Hunan Key Laboratory of Super-Microstructure and Ultrafast Process, School of Physics, Central South University, Changsha 410083, China}
\affiliation{Hunan Provincial Key Laboratory of Flexible Electronic Materials Genome Engineering, School
of Physics and Electronic Science, Changsha University of Science and Technology, Changsha 410114, China}
\author{Yu Guo}
\affiliation{Hunan Provincial Key Laboratory of Flexible Electronic Materials Genome Engineering, School
of Physics and Electronic Science, Changsha University of Science and Technology, Changsha 410114, China}
\author{Shan Ma} 
\affiliation{School of Automation, Central South University,
Changsha 410083, China}
\author{Chuan-Cun Shu}
\email{cc.shu@csu.edu.cn}
\affiliation{Hunan Key Laboratory of Nanophotonics and Devices, Hunan Key Laboratory of Super-Microstructure and Ultrafast Process, School of Physics, Central South University, Changsha 410083, China}

\begin{abstract}
We present a theoretical investigation of coherent control over the orientation of an individual molecule strongly coupled with a cavity using chirped-pulse driving. Specifically, we explore the dynamics of carbonyl sulfide (OCS) molecules under the influence of two chirped pulses with different spectral phases. We compare two pulse configurations: one with equal chirp rates ($\beta_{+} = \beta_{-}$) and another with unequal chirp rates ($\beta_{+} \neq \beta_{-}$). Numerical simulations reveal that chirped pulses enable precise control of the molecular orientation, achieving a maximum orientation degree of 0.5773. By analyzing the distribution of molecular polariton states, we show that chirped pulses can activate multiphoton processes, leading to deviations from the predictions of first-order Magnus expansion methods. Additionally, we demonstrate the robustness of the maximum orientation with respect to chirp amplitude and detuning, providing insights into the role of pulse parameters in optimizing control. This work introduces a new strategy for controlling molecular orientation in cavity-based systems and offers valuable perspectives for future experimental applications.

\end{abstract}
\maketitle
\section{Introduction}
Cavity quantum electrodynamics (QED) provides a fundamental framework for studying strong light–matter interactions, especially the formation of hybrid quasiparticles. When the interaction strength between quantum systems surpasses their combined decay and decoherence rates   \cite{hutchison2012modifying,garcia2021manipulating,nagarajan2021chemistry}, hybrid excitations known as polaritons emerge. Although early research focused on atomic systems, strong coupling between optical cavities and molecules now enables the creation of molecular polaritons that integrate photonic and molecular properties. These states have advanced charge and excitation transport   \cite{coles2014polariton,zhong2017energy,akulov2018long,semenov2019electron,krainova2020polaron,nagarajan2020conductivity,wang2021polariton}, tunable chemical reactivity   \cite{thomas2016ground,thomas2019tilting,lather2019cavity,vergauwe2019modification,ahn2023modification}, and new approaches to quantum state control. Ongoing research in cavity QED continues to deepen our understanding of light–matter coupling and drives innovation in quantum information, photonics, and molecular engineering.\\ \indent 
The Jaynes–Cummings (JC) model, introduced by E.T. Jaynes and F.W. Cummings in 1963, is a foundational tool for describing strong coupling between a two-level quantum system and a single electromagnetic field mode   \cite{jaynes1963comparison}. In cavity–molecule interactions, the JC model captures quantum superpositions of molecular and field states   \cite{ebbesen2016hybrid,ribeiro2018polariton}, known as polaritons, which support phenomena such as Rabi oscillations   \cite{brune1996quantum,blais2021circuit,kaiser2022cavity}, collapses and revivals   \cite{gea1990collapse,hunanyan2024periodic,redchenko2025observation}, photon blockade   \cite{zou2020multiphoton,li2024exploring}, and quantum phase transitions   \cite{greentree2006quantum,liu2024quantum}. Because real molecules have complex electronic, vibrational, and rotational structures, the JC model is often extended to account for them. These extensions enable detailed studies of strong coupling involving various molecular states, supporting investigations of ionization spectra, chemical reaction rates, and voltage-controlled switching   \cite{coulson2013electrically,hertzog2017voltage,xiang2019manipulating,galego2019cavity,gu2023manipulating,xiang2024molecular,mondal2025switching}. 
To investigate quantum control of molecular rotation within a cavity, we recently developed a pulse-driven JC model for a single-mode cavity strongly coupled to two rotational states of an ultracold molecule   \cite{fan2023L,fan2023}. We derived analytical solutions for frequency-domain control pulses. By optimizing the amplitude and phase of two pulses, the molecule’s maximum orientation inside the cavity matches that in free space. Time–frequency analysis reveals that the spectral phase distribution governs the relative phases among pulse frequency components   \cite{guo2022cyclic,hong2023quantum,fan2023L,fan2023}. Modulating the spectral phase enables temporal pulse shaping without altering the total fluence of the pulse. This method is effective for studying quantum coherence and interference, though  applying chirp rates changes the pulse peak intensity \cite{assion1998control,daniel2003deciphering,wilma2018visualizing,guo2019vanishing,nandi2022observation,jing2023unveiling,liu2024unveiling,richter2024strong,meron2025shaping,yang2025spectral}. \\ \indent 
To further examine the influence of spectral phase modulation, we extend the model in this work to include the spectral phase-modulated pulses into the JC model. We present a theoretical study of coherent control over the orientation of a single carbonyl sulfide (OCS) molecule strongly coupled to a cavity using chirped-pulse driving. We examine orientation dynamics induced by two chirped pulses with different spectral phases, comparing equal ($\beta_{+} = \beta_{-}$) and unequal ($\beta_{+} \neq \beta_{-}$) chirp rates. Numerical simulations show that chirped pulses enable precise and robust molecular orientation, reaching a maximum degree of 0.5773. Analysis of molecular polariton-state distributions reveals that chirped pulses can activate multiphoton processes, leading to deviations from predictions of the first-order Magnus expansion. The maximum orientation remains robust against changes in the chirp pulse’s amplitude and detuning, underscoring the importance of pulse parameters for optimal control. This work provides a chirped-pulse-driven JC model for molecular orientation control in cavity-based systems and highlights the broader potential of pulse shaping in quantum control applications.\\ \indent 
The remainder of this paper is organized as follows: Section \ref{Sec:Model} details the theoretical framework of the chirped-driven JC model. Section \ref{RESULTS AND DISCUSSION} presents numerical simulations and discusses the control schemes for both equal and unequal chirp rates. Finally, Section \ref{conclusion} summarizes the main findings.\\ \indent
\section{Theoretical Methods \label{Sec:Model}}
\begin{figure}[H]  
\centering
\includegraphics[width=0.5\textwidth, keepaspectratio]{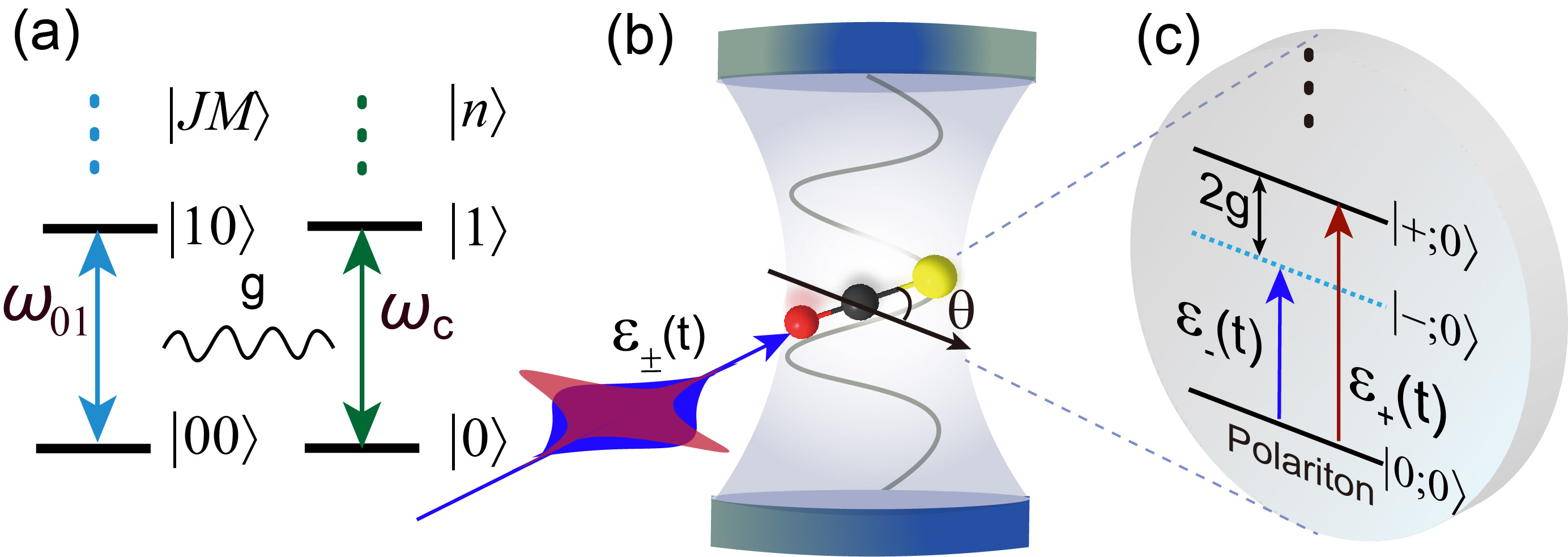}
\caption{Schematic illustration of a single molecule in a single-mode cavity driven by two chirped pulses (b), and the corresponding energy level diagrams in the bare eigenbasis (a) and the dressed basis (c).}
\label{fig1}
\end{figure}
As illustrated in Fig.\ref{fig1}, our model considers a single molecule within a single-mode cavity, characterized by a frequency $\omega_c$, that resonates with the lowest two rotational states of the molecule with a rotational constant $B$.  The Hamiltonian of the cavity-molecule system reads
\begin{equation}\label{hmt}
\hat{H}_{0}=B\hat{J}^{2}+\omega_{c}\hat{a}^{\dagger}{\hat{a}}-\sqrt{\frac{\omega_{c}}{2\epsilon_{0}V}}\hat{\mu}\cdot{\hat{\mathrm{e}}}({\hat{a}}+{\hat{a}}^{\dagger}),
\end{equation}
where $\hat{J}$ is the angular momentum operator, and $\hat{a}$ ($\hat{a}^{\dagger}$) represents the photon annihilation (creation) operator. The final term captures the interactions between the molecule and the cavity field, with $\epsilon_{0}$ denoting the electric constant, $V$ the volume of the electromagnetic mode, and $\hat{\mu}$ the molecular dipole operator. In Eq.~(1), we neglect the quadratic contributions, namely the cavity-field self-energy term
$\frac{g_0^{2}}{2}(\hat a+\hat a^\dagger)^2$
 and the molecular self-dipole-energy term
$\frac{g_0^{2}}{\omega_c}\hat{\mu}^{2}$
, which are known to be important for ensuring the stability (boundedness)
of the Hamiltonian and for preserving gauge consistency under matter-state truncation, especially in the ultrastrong-coupling regime  \cite{cohen1998atom,flick2017atoms,rokaj2018light,schäfer2020relevance,szidarovszky2023efficient}. In the present work, we consider the coupling strength $g_0/\omega_c \lesssim 0.1$, for which our previous study has demonstrated that these quadratic terms have a negligible impact and can therefore be safely neglected  \cite{fan2025maximizing}.

As we demonstrated in our previous work  \cite{fan2025maximizing}, by accounting for the strong coupling between the two low-lying rotational states of the molecule and the cavity, the Hamiltonian of the cavity-molecule (polariton) system can be represented using the JC model, and the corresponding Hamiltonian can be given by 
\begin{equation}\label{hmt}
\hat{H}_{JC}=\omega_{01}\vert 10\rangle\langle 10 \vert+\omega_{c}\hat{a}^{\dagger}{\hat{a}}+g({\hat{a}}\vert 10\rangle\langle 00 \vert+{\vert 00\rangle\langle 10 \vert\hat{a}}^{\dagger}),
\end{equation}
where $\omega_{01}$ represents the transition
frequencies of the molecular system, the coupling strength $g={\sqrt{\omega_{c}/(2\epsilon_{0}V)}}\langle00|\mu\cos\theta|10\rangle$ is significantly smaller than the frequency $\omega_{c}$. By diagonalizing the  Hamiltonian $\hat{H}_{\mathrm{JC}}$, we obtain the eigenvalues of the molecular polariton
%at the resonant cavity-molecular coupling $\omega_c=\omega_{01}$
\begin{eqnarray}
\omega_{0,0}&=&0,\nonumber\\
\omega_{\pm,n}&=&\omega_c(n+1)\pm g\sqrt{n+1}
\end{eqnarray}
and the corresponding eigenstates
\begin{eqnarray}
|0;0\rangle&=&|00\rangle|0\rangle,\nonumber\\
|\pm;n\rangle&=&\sqrt{2}/2\left(|00\rangle|n+1\rangle\pm|10\rangle|n\rangle\right).
\end{eqnarray} 
These dressed states form a doublet within each fixed total-excitation manifold $N=n+1$, arising from coherent hybridization of the bare states $|00\rangle|n+1\rangle$ and $|10\rangle|n\rangle$. The resulting polaritonic splitting scales as $2g\sqrt{n+1}$ (reducing to the vacuum Rabi splitting $2g$ for $n=0$).

Building upon our previous works  \cite{fan2023L,fan2023}, we determined that the maximum degree of orientation for such a polariton system can be attained by constructing a desired coherent superposition of a three-state system, comprising the states $\vert 0;0 \rangle$ and $\vert \pm;0 \rangle$, through photon blockade of higher dressed states. 
\subsection{The maximum degree of orientation for a three-state model}
To obtain the maximum degree of orientation, we consider a desired wave function involving three states at the target time $t_f$, which can be expanded as
\begin{eqnarray} \label{dwf}
\vert\psi(t_f)\rangle&=&\sum_{\ell=0,\pm}C_{\ell,0}(t_f)e^{-i\omega_{\ell,0}t_f}\vert\ell;0\rangle, 
\end{eqnarray}
where $C_{\ell,0}$ are the complex coefficients for the lowest three states $\vert 0;0\rangle$ and $\vert \pm;0\rangle$. The orientation of the three-level system is given by
\begin{eqnarray}\label{cs}
\langle\cos\theta\rangle&=&2\sum_{\ell=\pm}\vert C
_{0,0}\vert\vert C_{\ell,0}\vert \cos(-\omega_{\ell,0}t+\phi_{\ell,0})M_{\ell,0},
\end{eqnarray}
with $\phi_{\ell,0}=\arg[C_{0,0}(t)]-\arg[C_{\ell,0}(t)]$ and $M_{\ell,0}=\langle 0;0 \vert\cos\theta\vert\ell;0\rangle$.

Using the method of Lagrange multipliers  \cite{shu2020a,hong2021a}, the maximum degree of orientation is given by
\begin{eqnarray}\label{max}
f(|C_{0,0}|,|C_{-,0}|,|C_{+,0}|) &=& \lambda=\sqrt{M_{-,0}^{2}+ M_{+,0}^{2}}
\end{eqnarray}
which reaches a maximum value of 0.5774,
with
\begin{eqnarray}\label{ac}
\vert C_{0,0}(t_f)\vert&=&\sqrt{2}/2,\\ \nonumber
\vert C_{-,0}(t_f)\vert&=&\vert C_{+,0}(t_f)\vert=1/2,
\end{eqnarray}
provided the following phase relation is satisfied
\begin{eqnarray}\label{pc}
\omega_{-,0}\phi_{+,0}-\omega_{+,0}\phi_{-,0}=g\pi+2gk\pi.
\end{eqnarray}This result demonstrates the optimal configuration for achieving the maximum orientation in a three-level quantum system. The interplay between the magnitudes and phases of the coefficients $C_{0,0}$, $C_{-,0}$, and $C_{+,0}$ is crucial in realizing this maximum. In practical terms, coherent control protocols aiming to reach this orientation must carefully control both the population distribution among the three states and their relative phases, as dictated by the constraints in Eqs.~\eqref{ac} and ~\eqref{pc}.
\subsection{A coherent control scheme with the use of two chirped pulses}
To generate such a coherent superposition of three states, we employ two chirped pulses $\mathcal{E}_{+}(t)$ and $\mathcal{E}_{-}(t)$, to drive the initial state $\vert 0;0\rangle$ transition to the polariton states $\vert +;0\rangle$ and $\vert -;0\rangle$, respectively. The corresponding Hamiltonian of the polariton driven by control fields $\mathcal{E}_{+}(t)$ and $\mathcal{E}_{-}(t)$ can be given by  
\begin{eqnarray} \label{Hp}
\hat{H}_{\mathrm{p}}(t)&=&\sum_{\ell=\pm,0}\omega_{\ell,0}\vert\ell;0\rangle\langle \ell;0\vert\nonumber \\
&&-\mathcal{E}_{+}(t)\tilde{\mu}_{+,0}\Big(\vert+;0\rangle\langle0;0|+|0;0\rangle\langle +;0\vert\Big) \nonumber \\
&&-\mathcal{E}_{-}(t)\tilde{\mu}_{-,0}\Big(\vert-;0\rangle\langle0;0\vert+\vert0;0\rangle\langle -;0\vert\Big),
\end{eqnarray}
where $\tilde{\mu}_{\pm,0}=\pm\frac{\sqrt{2}}{2}\mu_{01}$ denote the transition dipole moments between the ground state $\vert 0;0\rangle$ and the excited state $\vert \pm;0\rangle$ with $\mu_{01}=\langle 00\vert\mu\cos\theta\vert 10\rangle=\frac{\sqrt{3}}{3}\mu$. To gain an insight into the time-frequency relation, we design the two time-dependent control pulses from the frequency domain 
\begin{equation}\label{ew2}
\mathcal{E}_{\pm}(t)=\frac{1}{\pi}\text{Re}\left[\int_{0}^{\infty} A_{\pm}(\omega)\exp[{i\phi_{\pm}(\omega)}]e^{i\omega t}d\omega\right],
\end{equation}
where $A_{\pm}(\omega)$  and  $\phi_{\pm}(\omega)$ denotes the spectral amplitudes and a spectral phases.  We consider the spectral amplitudes with Gaussian frequency distributions 
\begin{eqnarray} \label{GA}
A_{\pm}(\omega)=\mathcal{A}_{\pm}\exp\left[-\frac{(\omega-\omega_{\pm})^{2}\tau_{0}^{2}}{2}\right],
\end{eqnarray}
with the central frequency $\omega_{\pm}$, the pulse duration $\tau_{0}$, and the amplitude $\mathcal{A}_{\pm}$. By expanding the spectral phases as 
\begin{eqnarray}\label{Tl}
\phi_{\pm}(\omega_{\pm})=\varphi_{\pm}+\frac{1}{2}(\omega-\omega_{\pm})^{2}\beta_{\pm},
\end{eqnarray}
we can obtain two time-dependent control pulses
\begin{align}\label{et}
\mathcal{E}_{\pm}(t)
&=\sqrt{\frac{2}{\pi}}\frac{\mathcal{A}_{\pm}}{\tau_0}\,
\text{Re}\Bigg[ \sqrt{\frac{\tau_{0}^{2}}{\tau_{0}^{2}-i\beta_{\pm}}}\,
\nonumber\\
&\times\exp\Bigg(
-\frac{t^{2}}{2\tau_{\pm}^{2}} -\frac{i\beta'_{\pm}t^{2}}{2} +i\omega_{\pm}t +i\varphi_{\pm}
\Bigg) \Bigg],
\end{align}
where $\varphi_\pm$ and $\beta_{\pm}$ denote the absolute phases and chirp rates, and the temporal durations and chirp rates are $\tau_{\pm}^{2}=\tau_{0}^{2}(1+\beta_{\pm}^{2}/{\tau_{0}^{4}})$,
and $\beta'_{\pm}=\beta_{\pm}/(\tau_{0}^{4}+\beta_{\pm}^{2})$. Thus,  introducing the chirp rate $\beta_0$ in the spectral phases $\phi_{\pm}(\omega)$  changes the peak amplitudes, durations, and instantaneous frequencies of the time-dependent electric fields $\mathcal{E}_{\pm}(t)$. However, according to Parseval’s theorem, the total pulse energy, proportional to $\int |\mathcal{E}_{\pm}(t)|^2 dt$, is conserved under any purely dispersive (all-pass) spectral phase modulation, as $|\tilde{\mathcal{E}}_{\pm}(\omega)|=A_{\pm}(\omega)$ is unaffected.

\subsection{An analytical wave function and the optimal conditions}
We derive an analytical wave function for the three states to determine the optimal parameters of two chirped pulses that can maximize the degree of orientation at \( t_f \). The time-dependent wave function of the system from the initial time \( t_0 \) can be obtained by numerically integrating the unitary evolution operator
\begin{equation}
\hat{U}(t,t_{0})=\mathcal{I}-i\int_{t_{0}}^{t} d t^{\prime}\hat{H}_{I}(t^{\prime})\hat{U}(t^{\prime},t_{0}) 
\end{equation}
where $\hat{H}_{I}(t)=\text{exp}(i\hat{H}_{s}t)[-\sum_{\ell=\pm}\tilde{\mu}_{\ell,0}\mathcal{E}_{\ell}(t)]\text{exp}(-i\hat{H}_{s}t)$, with the field-free Hamiltonian of the three-level system $\hat{H}_{s}=\omega_{-,0}\vert -;0\rangle\langle-;0\vert+\omega_{+,0}\vert +;0\rangle\langle +;0\vert$.

We apply the  Magnus expansion to the unitary operator
\begin{eqnarray}
  \hat{U}(t,t_{0}) =\exp{\left[\sum_{n=1}^{\infty }\hat{S}^{(n)}(t)\right]},  
\end{eqnarray}
where the first three leading terms read
 $\hat{S}^{(1)}(t)= -i\int_{t_{0}}^{t}dt_{1}H_{I}(t_{1})$, $\hat{S}^{(2)}(t)= (-i)^{2}/2\int_{0}^{t}dt_{1}\int_{0}^{t_{1}}dt_{2}[H_{I}(t_{1}),H_{I}(t_{0})]$, $\hat{S}^{(3)}(t)= (-i)^{3}/6\int_{0}^{t}dt_{1}\int_{0}^{t_{1}}dt_{2}\int_{0}^{t_{2}}dt_{3}\{H_{I}(t_{1}),[H_{I}(t_{2}), H_{I}(t_{2})]\}$. By considering the first-order Magnus expansion and the system initially in the ground state $\vert 0;0\rangle$, the analytical time-dependent wave function for the three states in the interaction picture is given below
\begin{equation}\label{psir}
\begin{aligned}
\vert\psi(t)\rangle_{I}
&=\cos\theta_{0}(t)\vert 0;0\rangle
+i\frac{\theta_{-}^{*}(t)}{\theta_{0}(t)}\sin\theta_{0}(t)\vert -;0\rangle \\
&\quad +i\frac{\theta_{+}^{*}(t)}{\theta_{0}(t)}\sin\theta_{0}(t)\vert +;0\rangle,
\end{aligned}
\end{equation}
where $\theta_{0}(t)=\sqrt{\vert\theta_{-}(t)\vert^2+\vert\theta_{+}(t)\vert^2}$ with $\theta_{\pm}(t)=\tilde{\mu}_{\pm,0}\int_{t_{0}}^{t}dt^{\prime}\mathcal{E}_{\pm}(t^{\prime})\\e^{-i\omega_{\pm,0}t^{\prime}}$.

By comparing Eq. (\ref{psir}) with the desired wave function of the target states in Eq. (\ref{dwf}), the maximum degree of orientation in Eq.(\ref{cs}) occurs at

\begin{eqnarray}\label{cd2}
\vert\theta_{\pm}(t_{f})\vert=\frac{\sqrt{2}\pi}{8}+\frac{\sqrt{2}k\pi}{4}.
\end{eqnarray}
while satisfying the phase condition
\begin{eqnarray}\label{phsc}
&\omega_{-,0}\arg[\theta_{+}(t_{f})]-\omega_{+,0}\arg[\theta_{-}(t_{f})]\nonumber\\
&=2g\pi+2gk\pi, \ \ \ (k=0,\pm 1,\cdots).
\end{eqnarray}
In the case of  resonant excitations with \(\omega_{\pm}=\omega_{\pm,0}\), Eqs. (\ref{cd2}) and (\ref{phsc}) can be utilized to determine optimal amplitudes
\begin{eqnarray}\label{apc}
\mathcal{A}_{\pm}=\frac{\vert\theta_{\pm}\vert}{\tilde{\mu}_{\pm,0}},
\end{eqnarray}
and phases
\begin{eqnarray}\label{pcc}
\varphi_{\pm}=\arg[{\theta_{\pm}}]. 
\end{eqnarray}
By analyzing the values of amplitudes from Eqs. (\ref{GA}) and (\ref{Tl}), we can find  that the absolute value of the coefficients of three states in Eq. (\ref{psir}) are independent of the chirp rates \(\beta_{\pm}\) at the resonant condition, but their phases depend on the chirp rates. It is important to note that the analytical time-dependent wave function in Eq. (\ref{psir}) and the corresponding optimal parameters are valid only when excitation processes induced by the higher-order Magnus terms are negligible. In the following simulations, we will demonstrate scenarios in which chirp rates alter the populations of quantum states, suggesting that the model extends beyond the first-order Magnus expansion.\indent 
\section{RESULTS AND DISCUSSION   \label{RESULTS AND DISCUSSION}}
To examine how  chirp rates \(\beta_{\pm}\) affect the molecular orientation in the cavity, we use the OCS (carbonyl sulfide) molecule as an example in Fig.\ref{fig1}, which has a the rotational constant of \(B = 0.20286 \text{ cm}^{-1}\) and the dipole moment of \(\mu = 0.715 \) D. The time-dependent wave function \(|\Psi(n, \theta, t)\rangle\) of the cavity-molecule system is numerically obtained by considering a molecular-driven quantum optical model with the total Hamiltonian
\begin{equation}\label{hmt}
\hat{H}_{\text{tot}}(t)=\hat{H}_{0}-\hat{\mu}\cdot\hat{e}\left[\mathcal{E}_+(t)+\mathcal{E}_-(t)\right],
\end{equation}
without applying the first-order Magnus expansion. In our simulations, higher rotational states of \(J = 2,3,4\) and photon numbers of \(n=1, 2, 3\) are included. The degree of orientation of the molecule is calculated by \(\langle \cos\theta\rangle(t)=\sum_n\langle\Psi(n,\theta,t)|\cos\theta|\Psi(n, \theta, t)\rangle\).
\begin{figure}[htbp]
\centering{%
\includegraphics[width=1\linewidth]{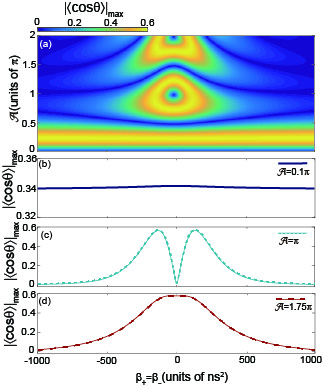}}\caption{Numerically calculated post-pulse orientation of the molecular polariton driven by chirped laser pulses with equal chirp rates $\beta_{+}=\beta_{-}$. 
(a) The maximum orientation values $\vert\langle\cos\theta\rangle\vert_{\text{max}}$ are shown vs the amplitude $\mathcal{A}_{\pm}$ and the chirp rates $\beta_{+}=\beta_{-}$. Cuts of $|\langle\cos\theta\rangle|_{\max}$ versus $\beta_{+}=\beta_{-}$ at three representative amplitudes: (b) $\mathcal{A}_{\pm}=0.1\pi$, (c) $\mathcal{A}_{\pm}=\pi$, and (d) $\mathcal{A}_{\pm}=1.75\pi$. Other parameters used in the numerical calculations are given in atomic units (a.u.): $\omega_-=1.66379\times10^{-6}\,\mathrm{a.u.}$, and $\omega_+=2.03353\times10^{-6}\,\mathrm{a.u.}$. The pulse duration $\tau_0=5.409\times10^{8}\,\mathrm{a.u.}\approx 13.0851\,\mathrm{ns}$, and the time interval is $t\in[-28\tau_0,\,28\tau_0]\approx[-366.383,\,366.383]\,\mathrm{ns}$.}
\label{fig2}
\end{figure}
\subsection{Simulations with equal chirp rates}
%To understand the impact of composite pulse chirp rates on the maximum orientation degree $\vert\langle\cos\theta\rangle\vert_{max}$ of molecular polaritons,
To resolve and selectively address the $+$ and $-$ polaritonic branches, the energy splitting must exceed both the effective polaritonic linewidth and the pulse bandwidth. Achieving high-fidelity state selectivity requires long cavity and molecular lifetimes, as well as a narrow chirped-pulse bandwidth. We first examine the case of two pulses with equal chirp rates $\beta_{+}=\beta_{-}$, with the pulse bandwidth set to $\Delta\omega=1/\tau_{0}=0.01g$ and relative phases fixed at $\phi_{-,0}=\pi$, $\phi_{+,0}=\pi/9$.\\ \indent 
Figure \ref{fig2} illustrates the dependence of $\vert\langle\cos\theta\rangle\vert_{\text{max}}$ on the pulse amplitudes $\mathcal{A}_{+}=\mathcal{A}_{-}=\mathcal{A}$ and the chirp rates $\beta_{\pm}$. The simulated parameter displays a striking landscape of alternating enhancement and suppression in the ($\mathcal{A}$, $\beta_{\pm}$) plane, providing evidence that the orientation of the molecular polariton is co-determined by pulse amplitude and chirp rate. The maximum orientation obtained from our simulation is highly sensitive to both parameters. For a given pulse amplitude $\mathcal{A}$, the effect of the chirp rate $\beta_{\pm}$ on the maximum orientation varies. For $\beta_{+}=\beta_{-}=0$(unchirped pulses),  increasing $\mathcal{A}$ produces a clear periodic modulation of $\vert\langle\cos\theta\rangle\vert_{\text{max}}$, consistent with the theoretical prediction from the first-order Magnus expansion. However, this periodicity is broken when $\beta_{\pm}\neq0$, the locations of maxima and minima shift, and the oscillatory structure becomes distorted. This behavior reflects the additional time-dependent detuning and dynamical phase imposed by the chirp, which alters the conditions for orientation accumulation compared with the unchirped case. \\ \indent To further explore this effect, Figs. \ref{fig2}(b)-(d) show the dependence on chirp rates at fixed amplitudes of $\mathcal{A}=0.1\pi$, $\pi$, and $1.75\pi$, respectively. Specifically, these figures plot $\vert\langle\cos\theta\rangle\vert_{\text{max}}$ as a function of $\beta_{\pm}$ at chosen $\mathcal{A}$, thereby highlighting how chirp sensitivity changes across intensity regimes. In the weak-amplitude ($\mathcal{A}=0.1\pi$), the maximum orientation shows negligible dependence on the chirp rate, with only slight deviations observed across a wide range of $\beta_{\pm}$. In contrast, in the medium/high-amplitude  ($\mathcal{A}=\pi$ and $1.75\pi$), $\vert\langle\cos\theta\rangle\vert_{\text{max}}$ becomes strongly chirp-dependent and exhibits pronounced optimal chirp windows, where the orientation is significantly enhanced to the theoretical upper bound of 
$0.5774$ in Figs. \ref{fig2}(c) and (d). These results demonstrate that the chirp rate provides an effective and practical knob for controlling molecular polariton orientation, particularly in regimes where the driving field is sufficiently strong for chirp-induced dynamical phases and detuning to play a dominant role.\\ \indent
\begin{figure*}[!t]
\centering
\includegraphics[width=0.9\textwidth]{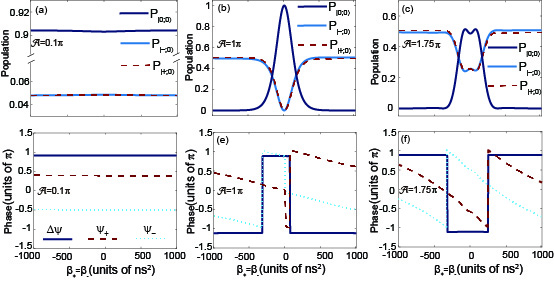}\caption{The final populations (a)-(c) and phases (d)-(f) of the three states $\vert 0;0\rangle$, $\vert \pm;0\rangle$ vs the chirp rate of the pusle at three different amplitudes  $\mathcal{A}=0.1\pi$, $\mathcal{A}=\pi$, and $\mathcal{A}=1.75\pi$. In panels (d)--(f), $\Psi_{+}$ and $\Psi_{-}$ denote the phases of the states $\lvert +;0\rangle$ and $\lvert -;0\rangle$, respectively, and $\Delta\psi$ is their phase difference. Note that the phases for larger detunings, where the populations of the states $\vert +;0\rangle$ and $\vert -;0\rangle$ are nearly zero, are meaningless and therefore are set to zero in our simulations.}
\label{fig3}
\end{figure*}
To gain insight into the physical mechanism by which the equal chirp rates affect the maximum orientation of molecular polaritons, Fig. \ref{fig3} presents the final-state populations [Figs. \ref{fig3}(a)-(c)] and phases [Figs. \ref{fig3}(d)-(f)] as functions of the equal chirp parameter $\beta{+}=\beta_{-}\equiv\beta$ for three pulse amplitudes $\mathcal{A}=0.1\pi$, $\pi$, and $1.75\pi$, corresponding to the orientation behavior in Figs. \ref{fig2}(b)-(d). This representation enables a direct correlation between chirp-induced redistribution among polariton states and the resulting changes in the attainable orientation. In the weak-amplitude ($\mathcal{A}=0.1\pi$), the final populations are essentially chirp-insensitive over the full range of $\beta_{\pm}$, indicating that the state composition remains nearly unchanged under spectral-phase variation. By contrast, at higher amplitudes ($\mathcal{A}=\pi$ and $\mathcal{A}=1.75\pi$), the populations become strongly chirp-dependent and exhibit pronounced transfer features, where population weight is redistributed among the polariton states as $\beta_{\pm}$ is varied. Interestingly, the pronounced chirp dependence of the final populations is not accounted for by the first-order Magnus approximation, indicating that higher-order contributions become non-negligible in the moderate-to-strong driving regimes.\\ \indent  For $\mathcal{A}=\pi$, increasing $\beta_{\pm}$ steers the system toward an orientation-favorable population partition, consistent with the enhanced orientation observed in Fig. \ref{fig2}. This correlation indicates that the improvement of orientation in this regime is primarily associated with chirp-controlled population transfer rather than a substantial chirp-induced modification of the final-state phases. For $\mathcal{A}=1.75\pi$, however, the redistribution pattern is qualitatively altered: the population transfer develops additional extrema (peak–valley features), and the final-state distribution deviates from the orientation-favorable (near-optimal) distribution. This behavior suggests that higher-order contributions in the Magnus expansion become significant in the strong-field regime and play an essential role in shaping the chirp-dependent population-transfer pathways. As shown in Figs. \ref{fig3}(d)-(f), the phase of the molecular polariton states is largely unaffected by the chirp rate, in line with theoretical predictions. Therefore, Fig. \ref{fig3} demonstrates that chirp control of molecular polariton orientation arises primarily from chirp-driven redistribution of polariton-state populations, while the phases remain comparatively insensitive over the explored chirp range, providing a clear mechanistic link between the spectral-phase parameters and the achievable orientation.\\ \indent 

\subsection{Simulations with  unequal chirp rates}
\begin{figure*}[!t]
\centering
\includegraphics[width=0.9\textwidth]{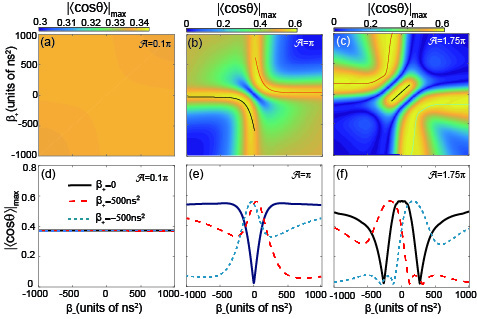}\caption{The maximum orientation values $\vert\langle\cos\theta\rangle\vert_{\text{max}}$ are shown vs two chirp rates $\beta_{+}$ and $\beta_{-}$ of the pulses at three different amplitudes (a) $\mathcal{A}=0.1\pi$, (b) $\mathcal{A}=\pi$, and (c) $\mathcal{A}=1.75\pi$. (e-f) Lower panels show cuts of $|\langle\cos\theta\rangle|_{\max}$ versus $\beta_{-}$ at fixed $\beta_{+}$, with $\beta_{+}=0$ (black solid), $\beta_{+}=+500~\mathrm{ns}^{2}$ (red dashed), and $\beta_{+}=-500~\mathrm{ns}^{2}$ (blue-dotted), corresponding to (a-c), respectively. All other parameters are the same as in Fig.~\ref{fig2}.}
\label{fig4}
\end{figure*}
We further examine the general case of unequal chirp rates($\beta_{-}\neq \beta_{+}$). Figure \ref{fig4} shows the maximum orientation $\vert\langle\cos\theta\rangle\vert_{\text{max}}$ as a function of the two chirp rates $\beta_{+}$ and $\beta_{-}$, for three different amplitudes: $\mathcal{A}=\mathcal{A_{+}}=\mathcal{A_{-}}=0.1\pi$, $\pi$ and $1.75\pi$, with other parameters remaining the same as in Fig. \ref{fig2}.
The orientation landscape is symmetric with respect to the origin, i.e., it is invariant under the simultaneous sign reversal $(\beta_{+},\beta_{-})\rightarrow (-\beta_{+},-\beta_{-})$. In the weak-field regime ($\mathcal{A}=0.1\pi$), $\vert\langle\cos\theta\rangle\vert_{\text{max}}$ depends only weakly on $\beta_{\pm}$. By contrast, for $\mathcal{A}=\pi$ and $\mathcal{A}=1.75\pi$, the maximum orientation becomes strongly dependent on both chirp rates, and the local maxima align along well-defined loci. To determine the optimal chirp-rate combinations that maximize the molecular polariton orientation, we numerically extract the loci of maximal $\vert\langle\cos\theta\rangle\vert_{\text{max}}$ in Figs. \ref{fig4}(b) and (c) and fit them with analytic functions. For $\mathcal{A}=\pi$, the fitted optimal relation is
\begin{eqnarray}
\beta_{+}=415\exp\left(\frac{-\beta_{-}}{91}\right)+32.5,  (\beta_{-}<0 \ \mathrm{ns}^2),\nonumber\\
\beta_{+}=-415\exp\left(\frac{\beta_{-}}{91}\right)+32.5,  (\beta_{-}>0 \ \mathrm{ns}^2),
\end{eqnarray}
where the two branches reflect the symmetry between positive and negative chirps.

For $\mathcal{A}=1.75\pi$, the optimal relation is well described by the piecewise form
\begin{eqnarray}
\beta_{+}&=&163+Y_{1}+Y_{2},  (-1000\ \mathrm{ns}^2)<\beta_{-}<-113 \ \mathrm{ns}^2),\nonumber\\
\beta_{+}&=&\beta_{-},  (-113\ \mathrm{ns}^2<\beta_{-}<113 \ \mathrm{ns}^2),\\
\beta_{+}&=&-(163+Y_{1}+Y_{2}),  (113\ \mathrm{ns}^2<\beta_{-}<1000 \ \mathrm{ns}^2),\nonumber
\end{eqnarray}
where $Y_{1}=157.44/(1+10^{0.0074(-321.6-\beta_{-})})$ and $Y_{2}=-1040.44/(1+10^{0.025(-167-\beta_{-})})$. It should be noted that the optimally fitted locus closely follows the diagonal when the chirp magnitude satisfies $(-113 \ \mathrm{ns}^2<\beta_{\pm}<113 \ \mathrm{ns}^2)$. As the chirp magnitude increases, the optimal locus becomes increasingly nonlinear, revealing a more complex interplay between the two chirp parameters in the regime.\\ \indent

To understand how the two chirp rates influence the maximum orientation of the molecular polariton, Figs. \ref{fig4}(e–f) displays the peak orientation as a function of the chirp rate $\beta_{-}$ for fixed values of $\beta_{+}$ ($-500$ ns$^{2}$, $0$, and $500$ ns$^{2}$). When $\beta_{+} = \beta_{-} = 0$, the maximum orientation matches the previously obtained unchirped theoretical value. For a low amplitude ($\mathcal{A} = 0.1\pi$), the chirp rate exerts minimal influence on the maximum orientation as $\beta_{-}$ increases. In contrast, at higher amplitudes ($\mathcal{A} = \pi$ and $\mathcal{A} = 1.75\pi$), not only does the chirp rate affect the magnitude of the maximum orientation, but its sign also determines the symmetry of the response. The orientation landscape is symmetric under the exchange $\beta_{-}\leftrightarrow\beta_{+}$, as shown by the diagonal reflection in Figs.~\ref{fig4}(a)-(c). It implies that swapping the two chirp rates does not change the orientation values. For $\mathcal{A}=\pi$, Fig.~\ref{fig4}(e) demonstrates that setting one chirp rate to zero and varying the other produces a broad plateau of near-maximum orientation. Thus, a single chirped pulse can achieve robust orientation in this regime, although robustness depends on the amplitude. Collectively, these findings demonstrate that as the pulse intensity increases, the chirp pair $(\beta_{+},\beta_{-})$ serves as a powerful control parameter, enabling precise steering of the system toward optimal regions in parameter space and thus maximizing molecular polariton orientation.\\ \indent 
\begin{figure*}[!t]
\centering
\includegraphics[width=0.90\textwidth]{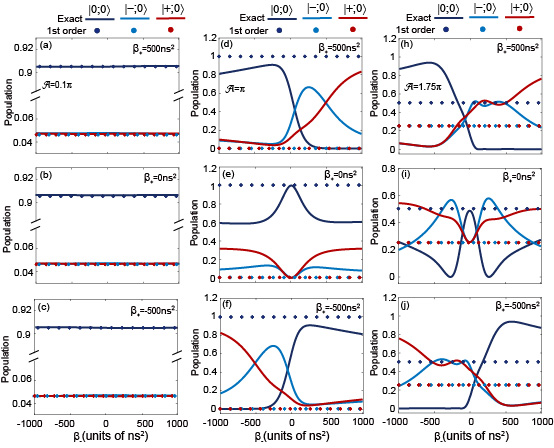}\caption{Final populations of the three polariton states versus the chirp rate $\beta_{-}$ for unequal-chirp driving, comparing exact numerical results with the first-order Magnus approximation. Panels (a)-(c), (d)-(f), and (h)-(j) correspond to three different amplitudes $\mathcal{A}_{\pm}=0.1\pi$, $\pi$, and $1.75\pi$, respectively. In each column, $\beta_{+}$ is fixed at (a,d,h) $500~\mathrm{ns}^{2}$, (b,e,i) $0$, and (c,f,j) $-500~\mathrm{ns}^{2}$, while $\beta_{-}$ is scanned. Solid curves show the exact final populations of $|0;0\rangle$ (black), $|-;0\rangle$ (blue), and $|+;0\rangle$ (red); filled circles indicate the corresponding first-order Magnus predictions. All other parameters are the same as in Fig.~\ref{fig2}.}
\label{fig5}
\end{figure*}
Figure \ref{fig5} shows the final populations of the molecular polariton states as functions of the chirp rate $\beta_{-}$ for three fixed values of $\beta_{+}$ ($\beta_{+} = -500$ $\ \mathrm{ns}^2$, $0$ $\ \mathrm{ns}^2$, and $500$ $\ \mathrm{ns}^2$). The exact numerical results are compared with the analytical predictions from the first-order Magnus approximation. Overall, the final populations exhibit a pronounced sensitivity to the chirp parameters, especially at moderate and strong amplitudes. In particular, when $\beta_{+} = \beta_{-} = 0$, the population distribution reduces to the unchirped result, consistent with the theoretical value. 
In the weak-field case $\mathcal{A} = 0.1\pi$ [Figs. \ref{fig5}(a)–\ref{fig5}(c)], the exact populations are nearly identical to the first-order Magnus prediction. However, at higher amplitudes ($\mathcal{A} = \pi$ [Figs. \ref{fig5}(d)–\ref{fig5}(f)] and $\mathcal{A} = 1.75\pi$ [Figs. \ref{fig5}(h)–\ref{fig5}(j)]), substantial discrepancies emerge between the exact results and the first-order approximation: the positions and magnitudes of population transfer features (peaks/plateaus and depletion regions) are shifted or reshaped, and the final-state composition can differ qualitatively. These deviations provide direct evidence that higher-order contributions in the Magnus expansion become significant in the strong-field regime and play an essential role in determining the chirp-dependent redistribution of polariton populations. Overall, Fig. \ref{fig5} clarifies the physical mechanism underlying chirp control: by modifying the time-dependent excitation dynamics, the chirp rates steer population transfer among polariton states, thereby setting the state distribution that ultimately governs the achievable molecular polariton orientation under chirped driving.
\subsection{Simulations with center-frequency- detuned pulses}
\begin{figure}[htbp]
\centering
\includegraphics[width=0.9\linewidth]{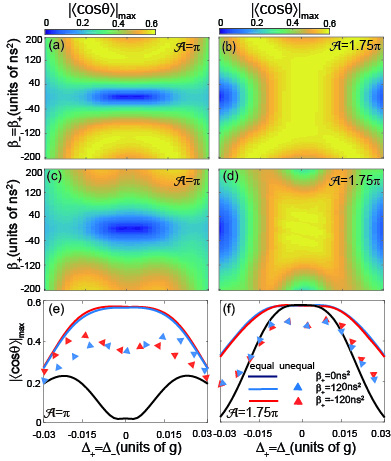}\caption{ The maximum orientation values
$|\langle\cos\theta\rangle|_{\max}$ are shown as functions of the equal chirp rate $\beta$ and the common detuning
$\Delta\equiv\Delta_{+}=\Delta_{-}$ for (a) $\mathcal{A}=\pi$  and  (b) $\mathcal{A}=1.75\pi$ .
Single-chirp configuration $\beta_{-}=0$: the maximum orientation values $|\langle\cos\theta\rangle|_{\max}$
are shown as functions of $\beta_{+}$ and $\Delta$ for  (c)$\mathcal{A}=\pi$ and  (d) $\mathcal{A}=1.75\pi$.
 Lower panels show cuts of $|\langle\cos\theta\rangle|_{\max}$ versus $\Delta_{+}=\Delta_{-}$ at fixed $\beta_{+}$: black solid curve,
unchirped case ($\beta=0$); red/blue solid curves, equal chirps at $\beta=\pm120~\mathrm{ns}^{2}$; triangles,
single-chirp case ($\beta_{-}=0$) at $\beta_{+}=\pm120~\mathrm{ns}^{2}$. All other parameters are the same as in
Fig.~\ref{fig2}.}
\label{fig6}
\end{figure}
To evaluate the experimental feasibility and robustness of our control scheme, we investigate how the maximum orientation responds to representative experimental imperfections. In particular, we consider (i) fluctuations in the chirp rates and (ii) detuning of the pulse center frequencies. Figure \ref{fig6} summarizes this robustness analysis by mapping the maximum orientation
$|\langle\cos\theta\rangle|_{\max}$ as a function of the common pulse-center-frequency detuning
$\Delta\equiv\Delta_{+}=\Delta_{-}$ and the chirp parameters, while keeping all other parameters identical to those in Fig.~\ref{fig2}. Two representative driving strengths are considered, $\mathcal{A}=\pi$ (left column) and $\mathcal{A}=1.75\pi$ (right column).  Figure \ref{fig6} (a) and (b) correspond to the equal-chirp configuration $\beta_{+}=\beta_{-}$, whereas \ref{fig6} (c) and (d) shows the single-chirp case with $\beta_{-}=0$ and variable $\beta_{+}$.\\ \indent
Figure \ref{fig6}  (a) and (b) display the equal-chirp configuration $\beta_{+}=\beta_{-}$, where $|\langle\cos\theta\rangle|_{\max}$ is mapped in the $(\beta,\Delta)$ plane.
Near the operating point $\beta=0$ and $\Delta=0$, the orientation reproduces the corresponding unchirped reference behavior, and high-orientation regions persist around small detuning. As the chirp rates increase, the maximum orientation follows the trend observed in Fig. \ref{fig2} (a). Specifically, the maximum orientation value shows a trend of first rising and then falling with $\mathcal{A}=\pi$, and the maximum orientation value gradually decreases when $\mathcal{A}=1.75\pi$. The results demonstrate that the maximum orientation remains near its optimal value over the detuning interval explored in Fig. \ref{fig6} around $\Delta_{\pm}=0$. Notably, the detuning dependence becomes weaker at larger chirp magnitudes (e.g. $\beta_{\pm} = \pm120 $ ns$^{2}$), indicating improved robustness against frequency detuning under strong chirping. Figure \ref{fig6}(c) and (d) present the single-chirp scenario ($\beta_{-}=0$), where $|\langle\cos\theta\rangle|_{\max}$ is plotted versus $\beta_{+}$ and $\Delta$.
Compared with the equal-chirp, the results demonstrate that the maximum orientation is strongly dependent on both the magnitude and the sign of the chirp rate $\beta_{+}$ and that the detuning response develops a more intricate pattern (with peak–valley features) than in the equal-chirp case. In particular, the maximum orientation exhibits an approximate symmetry under $\beta_{+}\rightarrow -\beta_{+}$ at fixed $\Delta_{+}=\Delta_{-}$, while its dependence on $\Delta_{+}=\Delta_{-}$ is noticeably less uniform. \\ \indent 
To further assess the impact of frequency detuning, we plot the corresponding cut lines of $\vert\cos\theta\vert_{max}$ versus frequency detuning $\Delta$ at fixed chirp rates $\beta_{\pm} = 0 $ ns$^{2}$ and  $\beta_{+} = \pm120 $ ns$^{2}$ for $\mathcal{A} = \pi$ in Fig. \ref{fig6} (e) and $\mathcal{A} = 1.75\pi$ in Fig. \ref{fig6} (f), respectively.  Figure \ref{fig6} (e) and (f) show representative cuts $|\langle\cos\theta\rangle|_{\max}(\Delta)$ extracted from the Figure \ref{fig6} (a)-(d). The black solid curve corresponds to the unchirped case. The red and blue solid curves show the equal-chirp configuration at $\beta_{\pm}=+120~\mathrm{ns}^{2}$ and $\beta_{\pm}=-120~\mathrm{ns}^{2}$, respectively, whereas the triangular markers represent the single-chirp scheme ($\beta_{-}=0$) evaluated at the corresponding $\beta_{+}=\pm120~\mathrm{ns}^{2}$.
A clear contrast emerges: the equal-chirp cuts (red/blue solid curves) exhibit a flatter, plateau-like dependence on $\Delta_{\pm}$, demonstrating improved tolerance to pulse-center-frequency detuning, whereas the single-chirp cuts (triangles) show stronger modulation and a more pronounced non-monotonic detuning response. As a result, the combined figure shows that applying equal chirps to both control pulses not only maintains a higher, more stable orientation near the optimal operating region but also enhances robustness against detuning, supporting the experimental feasibility of the equal-chirp control strategy. 
\section{conclusion\label{conclusion}}
We conducted a theoretical study on controlling molecular polariton orientation using phase modulation of composite chirped pulses. Numerical simulations for both equal- and unequal-chirp configurations allowed us to systematically examine how spectral phase influences maximum orientation and final polariton-state distribution. Our results indicated that chirp rates significantly affect molecular orientation. Comparative analysis showed that the phase dependence of maximum orientation mainly results from phase-induced nonresonant transitions. By comparing state distributions from the first-order Magnus expansion with exact solutions, we identified the limitations of the first-order approximation and clarified how chirped pulses induce nonresonant transitions in molecular polariton states. We also assessed the robustness of maximum orientation against pulse detuning. These findings offer valuable insights into controlling molecular polariton orientation and demonstrate the impact of spectral-phase modulation on polariton-state transitions, providing a chirped-pulse-driven JC model to explore the strong coupling of a rotational molecule with a cavity. \\ \indent 
Our findings demonstrate that adjusting field parameters, such as chirp rates, enables robust control over rotational dynamics in polariton systems. Future research could examine the effects of various pulse shapes, multi-mode coupling, or decoherence on these phenomena. While this work focuses on the lowest three states, the approach can be applied to systems with more levels, where optimization may require additional constraints and more complex phase relationships \cite{shu2016monotonic,shu2016frequency,van2017robust,koch2022quantum,hong2025precise}. Expanding these investigations will deepen understanding and control of molecular polariton systems, potentially enabling new quantum control and photonic applications through tailored light-matter interactions.
\section*{Acknowledgments}
  This work was supported by the National Natural Science Foundation of China (NSFC)  under Grants No.12274470, No. 12275033, and No.62273361. This work was carried out in part using computing resources at the High Performance Computing Center of Central South University.

%\bibliography{reference}
\end{document}